\documentclass[preliminary,copyright,creativecommons]{eptcs}
\providecommand{\event}{ICLP 2026 Technical Communication} 

\usepackage{hyperref}
\usepackage{iftex}

\ifpdf
 \usepackage{underscore}     
 \usepackage[T1]{fontenc}    
\else
 \usepackage{breakurl}      
\fi

\usepackage{amsmath}
\usepackage{graphicx}
\usepackage{multirow}
\usepackage{alltt}
\usepackage{amssymb}
\usepackage{tcolorbox}
\usepackage{esvect}
\usepackage{listings}
\usepackage{url}
\usepackage{plain}
\usepackage{float}
\usepackage{markdown}
\usepackage{booktabs}
\usepackage{rotating}

\title{Case study: solving P-99 with LPTP and an LLM}

\author{Fred Mesnard \qquad Thierry Marianne
\qquad \'Etienne Payet
\institute{LIM, universit\'e de La R\'eunion, France}
\email{\{frederic.mesnard,thierry.marianne,etienne.payet\}@univ-reunion.fr}
\and Wim Vanhoof
\institute{Universit\'e de Namur, Belgium}
\email{wim.vanhoof@unamur.be}
}

\newcommand{\ie}{\textit{i.e.}, }
\newcommand{\eg}{\textit{e.g.}, }

\newcommand{\titlerunning}{Case study: solving P-99 with LPTP and an LLM}
\newcommand{\authorrunning}{F. Mesnard, T. Marianne, \'E. Payet, W. Vanhoof}

\hypersetup{
 bookmarksnumbered,
 pdftitle  = {\titlerunning},
 pdfauthor  = {\authorrunning},
 pdfsubject = {EPTCS},        
 pdfkeywords = {Prolog, Verification, LLM} 
}

\begin{document}

\maketitle

\begin{abstract}

Ninety-Nine Prolog Problems (P-99) is a famous set of Prolog exercises.
We solved the first thirty three 
\emph{just by prompting an LLM} (Large Language Model). We used 
Claude from Anthropic. 
By ``solved'' we mean: generate the Prolog code and a test file,
run the tests and check whether they pass, then formally prove types, groundness,
termination, uniqueness, existence 
and also sometimes functional correctness with LPTP
(Logic Program Theorem Prover). 
Hence our approach is an experiment in \emph{vibe-coding/vericoding} of P-99.
It is a \emph{vibe-coding} experiment because we started from informal specifications written in English and let Claude generate the Prolog code.
It also fits within \emph{vericoding} because the LLM proved \emph{reliability guarantees} on the generated Prolog code. 
Claude wrote 58 logic procedures, 508 tests, 257 lemmas
for a total of 11800 proof lines.
We manually checked each file generated by the LLM.
We checked the Prolog code, ran the tests, examined the logical statements generated by Claude and proof-checked Claude's proofs with LPTP. 
This paper describes this experiment and provides the main details so that it can be reproduced by the interested reader.

\end{abstract}


\section{Introduction}
\label{Introduction}

Ninety-Nine Prolog Problems (P-99) is a famous set of Prolog exercises
that have been translated to many other programming languages,
including Erlang, Haskell, Lisp, Picat and Rust. 
There is also an Active Logic Document~\cite{MoralesAFH23}
for P-99 running Ciao-Prolog in any modern web browser\footnote{\url{https://cliplab.org/logalg/doc/99problemsALD.html}}. The oldest copy of P-99 we have dates back to more than 25 years. 
These exercises were written by Werner Hett from the Bern
University of Applied Sciences, Switzerland. 
The original site has been offline for many years.
There are a few copies floating on the Internet (\eg here\footnote{
\url{https://www.ic.unicamp.br/~meidanis/courses/mc336/2009s2/prolog/problemas/}}).
Some exercises are missing (P29, P30, P42-P45, P51-P53, P74-P79) and
a few of them have extensions (P61A, P62B, P70B, P70C). All in all, there are 88 exercises. 

In this work, we present a case study using the combination of an LLM and LPTP (Logic Program Theorem Prover)~\cite{Staerk98a} to solve some of the exercises of P-99. We choose Claude from Anthropic and do the experiment in \emph{Code mode} with the Opus 4.6 model released in February 2026.
We solve the first thirty three exercises (P01-P28 + P31-P35 = 33/88 = 37.5\%) 
just by \emph{prompting Claude}. By ``solve'' we mean: generate the Prolog code and a test file, run the tests and check whether they pass, then formally prove types, groundness, termination, uniqueness, existence and also sometimes functional correctness with LPTP.
In other words, our case study relates an experiment in \emph{vibe-coding/vericoding} of P-99. 
It is a \emph{vibe-coding}~\cite{sarkar2025vibecoding} experiment because we start from informal specifications written in English for students and let Claude generate the Prolog code.
It also fits within a \emph{vericoding} approach~\cite{DBLP:journals/corr/abs-2509-22908} because the LLM proved \emph{reliability guarantees} on the generated Prolog code in a formal language that can be proof-checked.

We did check manually each of Claude’s outputs. We checked the Prolog code, (re-)ran the tests, examined the logical statements generated by Claude and (re-)proof-checked Claude’s proofs with LPTP. For functional correctness properties (informally: properties stating that the code computes what it is supposed to), we almost always provided some hints to Claude to express the logical statements we were interested in. 

This paper describes this experiment and provides the main details so that it can be reproduced by the interested reader.
It is organized as follows. 
Section \ref{Goal} defines the experimental framework.
Sections~\ref{Claude-setup} and~\ref{Claude-md} 
describe how to instruct Claude about what it needs to know. 
Section \ref{Overview-of-the-experiment} gives an overview of the experiment.
Section \ref{Selected-examples} presents three examples with an emphasis
on functional correctness properties.
Section \ref{Current-work} describes an on-going work 
on implementing a Model Context Protocol connected to LPTP.
Section \ref{Related-work} reviews some related work 
and Section \ref{Conclusion} concludes.

\section{Goal}
\label{Goal}

The first problem of P-99 is given as follows:
\begin{verbatim}
  P01 (*) Find the last element of a list.
  Example:
  ?- my_last(X,[a,b,c,d]).
  X = d
\end{verbatim}

This is an informal specification written in English.
In general, a specification includes at least one example query that indicates the predicate name, arity and expected usage. 
The \texttt{(*)} indicates the difficulty of the exercise, ranging from one to three stars.

We want the LLM to propose the corresponding Prolog code
together with some tests. We want proven lemmas ensuring types, 
groundness, termination and functional properties (such as
links with library predicates or previously solved exercises).
We want to repeat this exercise for the next 32 problems.

More precisely, the programming language we instruct Claude to use is pure Prolog
with equality \texttt{=/2} on finite trees, negation as failure, the if/then/else construct and \emph{no} builtins. 
This subset is dictated by the input object language that LPTP
can deal with (although the subset can be slightly enlarged for some builtins which can be logically approximated~\cite{Staerk98a}). 
Natural numbers are represented using Peano notation (\eg $2$ is noted as \texttt{s(s(0))}).
The Prolog engine is assumed to run unification
\emph{with} occurs-check. SWI-Prolog has a special flag for this mode:
\texttt{set\_prolog\_flag(occurs\_check,true)}.
Claude adds the corresponding directive at the beginning of each test file.
The specification language and proof checker is LPTP
\cite{Staerk95a,Staerk96a,Staerk98a}, see also the companion paper~\cite{MesnardPV26} in this volume for a quick summary.

\section{Claude setup}
\label{Claude-setup}
Let us download the GitHub 
repository\footnote{\url{https://github.com/FredMesnard/LPTP-LLM-P99}}
accompanying this paper.
We get the \texttt{LPTP-LLM-P99-main} directory, which
contains the files \texttt{CLAUDE.md} (see the next section),
\texttt{P-99.html}, the P-99 Prolog problems in HTML format
and the file \texttt{lptp-reference.md}.
In this last file, Claude \emph{itself} maintains tips and lessons learnt 
about the project, such as 
how to run LPTP and Prolog locally,
common pitfalls and proof patterns. 

For each \texttt{Pxx} from \texttt{P99}, we want Claude 
to create a directory \texttt{P99/Pxx},
with its own \texttt{CLAUDE.md} file describing the problem, 
its solution \texttt{pxx.pl}, its tests \texttt{pxx\_test.pl},
the properties and their proofs \texttt{pxx.pr}, and 
a report \texttt{pxx-experiment-report.md} describing the work done.
The file \texttt{pxx.gr} is a representation created by Claude 
of the Prolog file \texttt{pxx.pl}, to be used by LPTP. 
Table~\ref{tab:repo-layout} displays the structure of the 
initial directory.

\begin{table}[H]
\centering
\caption{Layout of the GitHub repository accompanying the paper}
\label{tab:repo-layout}
\begin{tabular}{|l|l|}
\hline
\textbf{Directory} & \textbf{Content} \\
\hline
\texttt{P99/}     & \texttt{CLAUDE.md}, \texttt{P-99.html}, \texttt{lptp-reference.md} \\
\hline
\texttt{P99/P01/} & \texttt{CLAUDE.md}, \texttt{p01-experiment-report.md}, \texttt{p01.gr}, \texttt{p01.pl}, \\
                  & \texttt{p01.pr}, \texttt{p01\_test.pl} \\
\hline
\end{tabular}
\end{table}

The next section is an exact copy of the contents of the file 
\texttt{P99/CLAUDE.md}.
It contains all the instructions
for Claude to solve a P-99 exercise.
It has to be adapted to the local configuration.
Once all the infrastructure is ready,
the following prompt can be used for solving P02: 
\emph{Read CLAUDE.md,
lptp-reference.md, and solve the P02 exercise}.


\section{CLAUDE.md}
\label{Claude-md}

\subsection{Project}
Formal verification of the 99 Prolog Problems using LPTP
(Logic Program Theorem Prover) v1.06. The goal is to provide
Prolog code, a test file, and LPTP proofs of properties for
some of the 99 Prolog Problems.

\subsection{Reference}
Read \verb+/Users/fred/Desktop/P99/lptp-reference.md+
before writing or debugging any \verb+.pr+ file.
Update this reference if a new proof tip has been found.

\subsection{Structure}
The problems are listed in \verb+P-99.html+.
Each problem lives in a \verb+Pxx/+ directory containing:
\begin{itemize}
\itemsep-0.2em
\item \verb+pXX.pl+ — Prolog program
\item \verb+pXX.gr+ — ground representation (variable-free clause encoding for LPTP)
\item \verb+pXX.pr+ — proof file (lemmas, theorems)
\item \verb+pXX_test.pl+ — SWI-Prolog test file (see Testing below)
\item \verb+pXX-experiment-report.md+ — experiment report
\item \verb+CLAUDE.md+ — problem specification
\end{itemize}

\subsection{Conventions}
\begin{itemize}
\itemsep-0.2em
\item \textbf{Prolog} strictly ISO-compatible,
\verb+=+ (equality, \ie finite-tree unification),
\verb-\+- (negation as failure), and \verb+( ... -> ... ; ...)+
(the \verb+if-then-else+ construct) are allowed, but no cut,
no other \mbox{built-in}.
\item Use predicates from the LPTP lib when available, as it will help for the proofs.
\item Peano naturals: \verb+0+, \verb+s(0)+, \verb+s(s(0))+, \dots{} using the LPTP nat library.
\item Hierarchical lemma names: \verb+predicate:property+ or \verb+predicate:property:variant+.
\item The predicates defined in \verb+pXX.pl+ and \verb+pXX.gr+ must correspond exactly.
\item Copy \verb+.gr+ and \verb+.pr+ to \verb+/Users/fred/lptp/tmp/+ before verification.
\item Command: \verb+cd /Users/fred/lptp && printf "io__exec_file('tmp/pXX.pr').\nhalt.\n"+
\verb+|+ \verb+bin/lptp 2>&1+
\item If \verb+pXX.pr+ uses lemmas from another problem
(\eg P35 imports P31), declare the dependency via
\verb+:- needs_gr+ / \verb+:- needs_thm+ and document it in the local \verb+CLAUDE.md+.
\end{itemize}

\subsection{Testing}
For each problem Pxx, create a test file \verb+pXX_test.pl+ that:
\begin{itemize}
\itemsep-0.2em
\item Starts with \verb+:- set_prolog_flag(occurs_check,true).+ as the very first line.
\item Loads the program via \verb+:- [pXX].+
\item Includes helpers for Peano conversion (\verb+to_peano/2+, \verb+from_peano/2+, etc.) when needed.
\item Tests each predicate defined in \verb+pXX.pl+ with representative cases:
typical inputs, edge cases (empty list, 0, single element), and expected failures.
\item Tests each semantic property proved in \verb+pXX.pr+
(types, uniqueness, ordered, product, etc.) as a runtime check.
\item Prints \verb+OK+ or \verb+FAIL+ for each test case.
\item Ends with \verb+:- halt.+
\item Runs with: \verb+/Applications/SWI-Prolog.app/Contents/MacOS/swipl pXX_test.pl+
\end{itemize}

\subsection{Properties to verify systematically}
\begin{enumerate}
\itemsep-0.2em
\item \textbf{Types} (\verb+pred:types+) — type preservation (list, nat, etc.)
\item \textbf{Groundness} (\verb+pred:ground+) — ground inputs $\Rightarrow$ ground outputs
\item \textbf{Termination} (\verb+pred:termination+) — termination under preconditions
\item \textbf{Uniqueness} (\verb+pred:uniqueness+) — determinism of the result
\item \textbf{Existence} (\verb+pred:existence+) — existence of the result
\item \textbf{Functional Correctness} — any link with library predicates or previously solved exercises
\end{enumerate}

\subsection{Experiment Report}
Each \verb+pXX-experiment-report.md+ should contain:
\begin{itemize}
\itemsep-0.2em
\item \textbf{Problem statement} — what the predicate does
\item \textbf{Prolog code} — summary of the predicates defined
\item \textbf{Properties proved} — list of lemmas with their statement and
proof technique (completion, structural induction, strengthened induction, algebraic, etc.)
\item \textbf{Statistics} — line counts (\verb+.pl+, \verb+.gr+, \verb+.pr+),
number of lemmas, proof-to-code ratio
\item \textbf{Difficulties and lessons learned} — LPTP pitfalls encountered, proof strategies that worked
\end{itemize}

\subsection{Language}
\begin{itemize}
\itemsep-0.2em
\item Code, lemma names, and LPTP proofs: English.
\item Experiment reports and documentation: English.
\item Conversation with the user: French (the user's preferred language).
\end{itemize}

\section{Overview of the experiment}
\label{Overview-of-the-experiment}

Together with Claude, we solved the first 33 Prolog exercises.
Understanding the specification, writing the Prolog code
and the test file takes a few minutes.
We noticed that from time to time,
a \emph{context compaction} is followed by 
\emph{this session is being continued from a previous conversation that ran out of context}. 
It can be useful to tell Claude to reread
the \texttt{CLAUDE.md} files as Claude 
may have forgotten some guidelines.
Switching to the verification part, 
the process is much slower.
For functional properties, 
we had to give hints to Claude.
For instance, for P01 and P02,
we asked: \emph{what is the connection with append/3?} Sometimes we had to fully formulate the properties
in natural language (see Section~\ref{P35}).
Solving one exercise takes Claude between 15 minutes (\eg P01) to several hours (\eg P35). We checked manually each generated 
file.
We checked the Prolog code, we ran the tests, we examined the logical statements and we checked the proofs with LPTP. 

During this experiment, Claude created 58 logic procedures, with a total of 112 Prolog clauses, 150~lines of code and 508 runtime tests. 
It did make use of negation as failure
but made no use of the if/then/else construct.
Although P-99 is a well-known Prolog resource,
as we explicitly asked for pure Prolog code, the code generated by Claude is often quite different from the Prolog solutions one may find on the Internet (see \eg Section~\ref{P31} and Section~\ref{P35}), which very frequently include
impure constructs.
Claude proved 257 lemmas and wrote about 11800 lines of proof.
It had no problem in using what is 
already available in the LPTP library, code and lemmas. 
While proving its lemmas, it found various proof techniques 
which it added and documented in the \texttt{lptp-reference.md} file.
An example of such a proof technique is the following: 
sometimes we have to strengthen the property to be proved
so that we can do the inductive proof, and then weaken the 
property to get the original one.
The human readable \texttt{lptp-reference.md} file 
is maintained by Claude and is included in the GitHub repository 
accompanying this paper to speed up another run of this experiment.

\section{Selected examples}
\label{Selected-examples}

\subsection{P01: the last element of a list}
\label{P01}

The problem statement of exercise P01 was given in Section~\ref{Goal}.
Claude produces the following Prolog code (naive solution):
\begin{verbatim}
  my_last(X, [X]).
  my_last(X, [_|L]) :- my_last(X, L).
\end{verbatim}
It also infers and proves 10 properties, see Table \ref{tab-p01-properties}. We simplify the LPTP syntax for readability.
The full statements and their LPTP-checked proofs
are listed in the \texttt{p01.pr} file.
The \texttt{p01\_test.pl} file contains 21 tests, both for the source code and for the first 7 properties.
The last three properties are Claude's answer to our question: 
\emph{what is the connection with append/3?} Actually,
properties 8 and 10 constitute a characterization of \texttt{my\_last/2}
w.r.t. \texttt{append/3}.
\begin{table}[ht]
\centering
\caption{Properties proved for \texttt{my\_last/2}}
\label{tab-p01-properties}
\small
\begin{tabular}{|c|l|p{7.5cm}|}
\hline
\textbf{\#} & \textbf{Property} & \textbf{Statement} \\
\hline
1 & Termination
  & $\forall x,l.\; \mathtt{list}(l) \Rightarrow \mathbf{terminates}\; \mathtt{my\_last}(x,l)$ \\
2 & Type preservation
  & $\forall x,l.\; \mathtt{my\_last}(x,l) \Rightarrow \mathtt{list}(l)$ \\
3 & Groundness
  & $\forall x,l.\; \mathtt{my\_last}(x,l) \wedge \mathbf{gr}(l) \Rightarrow \mathbf{gr}(x)$ \\
4 & Membership
  & $\forall x,l.\; \mathtt{my\_last}(x,l) \Rightarrow \mathtt{member}(x,l)$ \\
5 & Uniqueness
  & $\forall x,l.\; \mathtt{my\_last}(x,l) \Rightarrow (\forall y.\; \mathtt{my\_last}(y,l) \Rightarrow x = y)$ \\
6 & Existence (aux.)
  & $\forall z,l.\; \mathtt{list}(l) \Rightarrow \exists x.\; \mathtt{my\_last}(x,[z|l])$ \\
7 & Existence
  & $\forall l.\; \mathtt{list}(l) \wedge l \neq [] \Rightarrow \exists x.\; \mathtt{my\_last}(x,l)$ \\
8 & Forward append
  & $\forall x,l.\; \mathtt{my\_last}(x,l) \Rightarrow \exists l_1.\; \mathtt{append}(l_1,[x],l)$ \\
9 & Converse append (aux.)
  & $\forall x,l_1,l.\; \mathtt{list}(l_1) \wedge \mathtt{append}(l_1,[x],l) \Rightarrow \mathtt{my\_last}(x,l)$ \\
10 & Converse append
  & $\forall x,l_1,l.\; \mathtt{append}(l_1,[x],l) \Rightarrow \mathtt{my\_last}(x,l)$ \\
\hline
\end{tabular}
\end{table}

\subsection{P31: a primality test}
\label{P31}

P31 asks for a primality test. 

\begin{verbatim}
  P31 (**) Determine whether a given integer number is prime.
  Example:
  ?- is_prime(7).
  Yes
\end{verbatim}
Claude generates the following Prolog code:

\begin{verbatim}
  % divides(D, N) - D divides N (by repeated subtraction, D > 0).
  divides(D, D).
  divides(D, N) :- plus(D, M, N), divides(D, M).

  % no_factor(K, D, N) - none of D, D+1, ..., D+K-1 divides N.
  %  K is a countdown: the number of candidates left to check.
  no_factor(0, _, _).
  no_factor(s(K), D, N) :- \+ divides(D, N), no_factor(K, s(D), N).

  % is_prime(N) - N is prime (N >= 2 and no factor from 2 to N-1).
  %  For N = s(s(X)), check X candidates starting from 2.
  is_prime(s(s(X))) :- no_factor(X, s(s(0)), s(s(X))).
\end{verbatim}

In Table~\ref{tab-p31-properties}, we extract the two 
properties\footnote{Numbering as in the \texttt{p31.pr} file.}
stating soundness and completeness of the implementation.
We did ask for such properties and gave the idea of their formalization in natural language. 
Then Claude was able to formalize
and prove them. 
\begin{table}[ht]
\centering
\caption{Soundness and completeness for \texttt{is\_prime/1}}
\label{tab-p31-properties}
\small
\begin{tabular}{|c|l|p{7.5cm}|} 
\hline
\textbf{\#} & \textbf{Property} & \textbf{Statement} \\
\hline
7 & Soundness of \texttt{is\_prime}
  & $\forall n.\; \mathtt{nat}(n) \wedge \mathtt{is\_prime}(n) \Rightarrow {}$ \newline $\exists x.\; n = s(s(x)) \wedge (\forall d_1.\; s(0) \leq d_1 \wedge d_1 < s(x) \Rightarrow $ \newline $\mathbf{fails}\; \mathtt{divides}(s(d_1), n))$ \\
9 & Completeness of \texttt{is\_prime}
  & $\forall n, x.\; \mathtt{nat}(n) \wedge n = s(s(x)) \wedge {}$ \newline $(\forall d_1.\; s(0) \leq d_1 \wedge d_1 < s(x) \Rightarrow$  \newline $\mathbf{fails}\; \mathtt{divides}(s(d_1), n))$ \newline $\Rightarrow \mathtt{is\_prime}(n)$ \\
\hline
\end{tabular}
\end{table}

\subsection{P35: decomposition in prime numbers}
\label{P35}

P35 asks for a decomposition of a natural number into prime factors.
\begin{verbatim}
  P35 (**) Determine the prime factors of a given positive integer.
  Construct a flat list containing the prime factors in ascending order.
  Example:
  ?- prime_factors(315, L).
  L = [3,3,5,7]
\end{verbatim}
Here is the Prolog code generated by Claude:
\begin{verbatim}
  % quot(D, N, Q) - quotient Q = N/D (assumes D divides N, D > 0).
  quot(D, D, s(0)).
  quot(D, N, s(Q)) :- plus(D, M, N), quot(D, M, Q).

  % smallest_factor(N, D, K, F) - smallest factor F of N starting from
  %  candidate D, with countdown K for termination.
  %  K counts how many candidates remain to try after D.
  smallest_factor(N, D, K, D) :- divides(D, N).
  smallest_factor(N, D, s(K), F) :- \+ divides(D, N),
    smallest_factor(N, s(D), K, F).

  % prime_factors(N, L) - L is the list of prime factors of N (>= 1)
  %  in ascending order, with repetitions.
  %  N = 0 fails (not a positive integer).
  prime_factors(s(0), []).
  prime_factors(s(s(X)), [F|L]) :-
    smallest_factor(s(s(X)), s(s(0)), X, F),
    quot(F, s(s(X)), Q),
    prime_factors(Q, L).
\end{verbatim}

Table \ref{tab-p35-properties} lists the properties 
of this program. Claude started with termination and type properties.
For stating and proving properties 23 to 30
that we explicitly asked in natural language,
Claude reused
code and lemmas from the \texttt{nat} library
(\eg \texttt{nat/1}, \texttt{plus/3}, \texttt{times/3}, 
\texttt{@=< /2}),
derived and proved new lemmas for P31 (\texttt{divides/2}),
P35 (\texttt{quot/3},
\texttt{smallest\_factor/4}, \texttt{prime\_factors/2})
and added Prolog code (\texttt{ordered/1}, \texttt{product/2}):
\begin{verbatim}
  % ordered(L) - L is sorted in ascending order (using @=<).
  ordered([]).
  ordered([_]).
  ordered([X,Y|L]) :- X @=< Y, ordered([Y|L]).

  % product(L, P) - P is the product of elements of L.
  product([], s(0)).
  product([X|L], P) :- product(L, P1), times(X, P1, P).
\end{verbatim}
We note that the product of the natural numbers contained in the empty list is 1.
Let us focus on the last few properties. Assuming
$n$ is a Peano integer and $prime\_factors(n,l)$ succeeds,
here is what Claude proved:
\begin{itemize}
\itemsep-0.2em
  \item property 23 (or 24): the resulting list $l$ 
is a list of Peano integers;
  \item property 26: the product of the elements of $l$ is $n$;
  \item property 27: $l$ is in ascending order;
  \item property 28: each element of $l$ is a prime number;
  \item property 30: $l$ is unique.
\end{itemize}
We also get a sufficient condition for the existence of a solution:
\begin{itemize}
  \item property 29: if $n$ is strictly greater than 0, there exists $l$ such that $prime\_factors(n,l)$ succeeds.
\end{itemize}
Actually the condition is also necessary as $prime\_factors(0,\_)$
finitely fails.
All together, these results constitute a logic-programming-based proof
of the \emph{prime factorization theorem}: 
``every integer greater than 1 is either prime or can be represented uniquely as a product of prime numbers, up to the order of the 
factors''\footnote{\url{https://en.wikipedia.org/wiki/Fundamental_theorem_of_arithmetic}}. Note that the Prolog code states that the prime factors of 1 is the empty list, which is meaningful.
So on the one hand, the Prolog code effectively constructs the ordered list $l$ of prime factors of a strictly positive natural number $n$. 
On the other hand, the proofs certify that if $n$ is strictly greater than 0 then $l$ always exists, is unique and correct.

\begin{sidewaystable}
\small
\begin{tabular}{|c|l|p{18cm}|}
\hline
\textbf{\#} & \textbf{Property} & \textbf{Statement} \\
\hline
\multicolumn{3}{|l|}{\textit{\texttt{quot/3} — quotient}} \\
1 & Termination
  & $\forall d_0,n,q.\; \mathtt{nat}(d_0) \wedge \mathtt{nat}(n) \Rightarrow \mathbf{terminates}\; \mathtt{quot}(s(d_0),n,q)$ \\
2 & Types
  & $\forall d_0,n,q.\; \mathtt{nat}(d_0) \wedge \mathtt{nat}(n) \wedge \mathtt{quot}(s(d_0),n,q) \Rightarrow \mathtt{nat}(q)$ \\
3 & Strict bound
  & $\forall d_1,n,q.\; \mathtt{nat}(d_1) \wedge \mathtt{nat}(n) \wedge \mathtt{quot}(s(s(d_1)),n,q) \Rightarrow q < n$ \\
4 & Correctness
  & $\forall d_0,n,q.\; \mathtt{nat}(d_0) \wedge \mathtt{nat}(n) \wedge \mathtt{quot}(s(d_0),n,q) \Rightarrow s(d_0) \times q = n$ \\
5 & Positive
  & $\forall d,n,q.\; \mathtt{quot}(d,n,q) \Rightarrow \exists q_0.\; q = s(q_0)$ \\
6 & Success
  & $\forall d,n.\; \mathtt{nat}(d) \wedge \mathtt{nat}(n) \wedge \mathtt{divides}(d,n) \Rightarrow \exists q.\; \mathtt{quot}(d,n,q)$ \\
7 & Uniqueness
  & $\forall d_0,n,q_1,q_2.\; \mathtt{nat}(d_0) \wedge \mathtt{nat}(n) \wedge \mathtt{quot}(s(d_0),n,q_1) \wedge \mathtt{quot}(s(d_0),n,q_2) \Rightarrow q_1 = q_2$ \\
\hline
\addlinespace
\hline
\multicolumn{3}{|l|}{\textit{\texttt{smallest\_factor/4}}} \\
8 & Termination
  & $\forall k,d_0,n,f.\; \mathtt{nat}(k) \wedge \mathtt{nat}(d_0) \wedge \mathtt{nat}(n) \Rightarrow \mathbf{terminates}\; \mathtt{smallest\_factor}(n,s(d_0),k,f)$ \\
9 & Types
  & $\forall k,d_0,n,f.\; \mathtt{nat}(k) \wedge \mathtt{nat}(d_0) \wedge \mathtt{nat}(n) \wedge \mathtt{smallest\_factor}(n,s(d_0),k,f) \Rightarrow \mathtt{nat}(f)$ \\
10 & Lower bound
  & $\forall k,d_0,n,f.\; \mathtt{nat}(k) \wedge \mathtt{smallest\_factor}(n,s(d_0),k,f) \Rightarrow \exists f_0.\; f = s(f_0)$ \\
11 & Lower bound 2
  & $\forall k,d_0,n,f.\; \mathtt{nat}(k) \wedge \mathtt{smallest\_factor}(n,s(s(d_0)),k,f) \Rightarrow \exists f_1.\; f = s(s(f_1))$ \\
12 & Divides
  & $\forall k,d_0,n,f.\; \mathtt{nat}(k) \wedge \mathtt{smallest\_factor}(n,s(d_0),k,f) \Rightarrow \mathtt{divides}(f,n)$ \\
13 & Completeness
  & $\forall k,d_0,n,\mathit{last}.\; \mathtt{nat}(k) \wedge \mathtt{nat}(d_0) \wedge \mathtt{nat}(n) \wedge \mathtt{plus}(d_0,k,\mathit{last}) \wedge \mathtt{divides}(s(\mathit{last}),n) \Rightarrow \exists f.\; \mathtt{smallest\_factor}(n,s(d_0),k,f)$ \\
14 & Leq factor
  & $\forall k,d_0,n,f,g.\; \mathtt{nat}(k) \wedge \mathtt{nat}(d_0) \wedge \mathtt{nat}(n) \wedge \mathtt{smallest\_factor}(n,s(d_0),k,f) \wedge \mathtt{divides}(g,n) \wedge s(d_0) \leq g \wedge \mathtt{nat}(g) \Rightarrow f \leq g$ \\
15 & Uniqueness
  & $\forall k,d_0,n,f_1,f_2.\; \mathtt{nat}(k) \wedge \mathtt{nat}(d_0) \wedge \mathtt{nat}(n) \wedge \mathtt{smallest\_factor}(n,s(d_0),k,f_1) \wedge \mathtt{smallest\_factor}(n,s(d_0),k,f_2) \Rightarrow f_1 = f_2$ \\
16 & Is prime
  & $\forall k,n,f.\; \mathtt{nat}(k) \wedge \mathtt{nat}(n) \wedge \mathtt{smallest\_factor}(n,s(s(0)),k,f) \Rightarrow \mathtt{is\_prime}(f)$ \\
\hline
\addlinespace
\hline
\multicolumn{3}{|l|}{\textit{\texttt{divides/2} — auxiliary}} \\
17 & Self
  & $\forall d.\; \mathtt{divides}(d,d)$ \\
18 & Sum
  & $\forall d,n,b,c.\; \mathtt{divides}(d,n) \wedge \mathtt{divides}(d,b) \wedge \mathtt{nat}(n) \wedge \mathtt{nat}(b) \wedge \mathtt{nat}(d) \wedge \mathtt{plus}(n,b,c) \Rightarrow \mathtt{divides}(d,c)$ \\
19 & Times factor
  & $\forall f_0,g,q.\; \mathtt{nat}(f_0) \wedge \mathtt{nat}(q) \wedge \mathtt{nat}(g) \wedge \mathtt{divides}(g,q) \Rightarrow \mathtt{divides}(g, s(f_0) \times q)$ \\
20 & Transitive
  & $\forall a,b,c.\; \mathtt{nat}(a) \wedge \mathtt{nat}(b) \wedge \mathtt{nat}(c) \wedge \mathtt{divides}(a,b) \wedge \mathtt{divides}(b,c) \Rightarrow \mathtt{divides}(a,c)$ \\
\hline
\addlinespace
\hline
\multicolumn{3}{|l|}{\textit{\texttt{prime\_factors/2}}} \\
21 & Termination
  & $\forall n,l.\; \mathtt{nat}(n) \Rightarrow \mathbf{terminates}\; \mathtt{prime\_factors}(s(n),l)$ \\
22 & Types
  & $\forall n,l.\; \mathtt{nat}(n) \wedge \mathtt{prime\_factors}(n,l) \Rightarrow \mathtt{list}(l)$ \\
23 & Nat members
  & $\forall n,l.\; \mathtt{nat}(n) \wedge \mathtt{prime\_factors}(n,l) \Rightarrow (\forall z.\; \mathtt{member}(z,l) \Rightarrow \mathtt{nat}(z))$ \\
24 & Nat list
  & $\forall n,l.\; \mathtt{nat}(n) \wedge \mathtt{prime\_factors}(n,l) \Rightarrow \mathtt{nat\_list}(l)$ \\
25 & Head info
  & $\forall n,h,t.\; \mathtt{nat}(n) \wedge \mathtt{prime\_factors}(n,[h|t]) \Rightarrow \mathtt{divides}(h,n) \wedge \mathtt{nat}(h) \wedge (\exists f_1.\; h = s(s(f_1)))$ \\
26 & Product
  & $\forall n,l.\; \mathtt{nat}(n) \wedge \mathtt{prime\_factors}(n,l) \Rightarrow \mathtt{product}(l,n)$ \\
27 & Ordered
  & $\forall n,l.\; \mathtt{nat}(n) \wedge \mathtt{prime\_factors}(n,l) \Rightarrow \mathtt{ordered}(l)$ \\
28 & All prime
  & $\forall n,l,z.\; \mathtt{nat}(n) \wedge \mathtt{prime\_factors}(n,l) \wedge \mathtt{member}(z,l) \Rightarrow \mathtt{is\_prime}(z)$ \\
29 & Existence
  & $\forall n.\; \mathtt{nat}(n) \Rightarrow \exists l.\; \mathtt{prime\_factors}(s(n),l)$ \\
30 & Uniqueness
  & $\forall n,l_1,l_2.\; \mathtt{nat}(n) \wedge \mathtt{prime\_factors}(n,l_1) \wedge \mathtt{prime\_factors}(n,l_2) \Rightarrow l_1 = l_2$ \\
\hline
\end{tabular}
\caption{Properties proved for P35}
\label{tab-p35-properties}
\end{sidewaystable}

\section{Proof generation and certification using an MCP}
\label{Current-work}

\subsection{MCP server tools}

In November 2024, Anthropic released the Model Context Protocol (MCP) \cite{mcpanthropic}. 
Relying on this open-source standard, we have implemented the \texttt{atp-lptp-mcp} server\footnote{\url{https://www.npmjs.com/package/atp-lptp-mcp}} by using the TypeScript SDK. It allows conversational assistants and agentic tools such as Claude Code to use LPTP directly. 
Concretely, the MCP server exposes LPTP-tailored tools 
to assistants like Claude Code and Gemini. The generative AI agent acts as an MCP client; it generates proof terms that are submitted to the MCP server and thus checked by LPTP.

As such, using an MCP server is a viable alternative to providing the \texttt{lptp-reference.md} file prior to generating the proof terms.
Using the MCP server offers several advantages. First, LPTP's ISO-Prolog syntax is embedded within the server via the \texttt{get\_lptp\_grammar} tool. Furthermore, 
as we no longer need to provide assistant-specific documentation in Markdown format 
or the LPTP reference, using MCP reduces token consumption during each interaction.
Finally, we can easily reuse the same MCP configuration with different LLM-based assistants and compare the outputs.

\subsection{Interactive Theorem Proving and LLM-based assistants}
Our MCP server maps calls to LPTP tactics such as induction, case analysis,
completion, definition unfolding, existential formula elimination, 
the totality axiom expansion or automated proof search.
These tactics are applied until a complete derivation is generated.
We opted for mapping each native tactic to a separate tool so that we
could request the LLM assistants for proof sketches containing tactic sequences,
before realizing that this intermediate step was not required.
The final architecture embeds the most general \texttt{apply\_tactic}
tool to apply arbitrary tactics depending on the target proof.
Regular proof checking is requested to LPTP via dedicated MCP tools. 

We instructed both Gemini 3.1 Pro and Claude Code (with its cheaper Sonnet 4.6 model) to synthesize LPTP properties (proofs with gaps) from logic programs corresponding to P14-P24 in P-99.
This subset is sufficiently diverse to illustrate properties such as
termination, determinism, groundness, type, existence and uniqueness. 
We noticed that Gemini generates a more diverse set of properties than Claude. 
The LLMs were then instructed to fill the gaps. 
We observed that only Claude generated valid proofs for the P14-P24 subset.

The MCP implementation can verify individual proof steps (by applying \texttt{verify\_lemma} to a single proof term) as well as entire proof files (\texttt{check\_proof}). It runs LPTP on top of SWI-Prolog 10.0.2. 
The MCP server converts the Prolog interpreter's answers to queries 
into JSON structures, again consumed by the LLM assistant. The LLM assistant keeps revising and submitting the produced derivations until they are certified by LPTP. 
This architecture ensures there are only two possible states at the end of the process: 
proofs with gaps or formally certified proofs. 
All property batches were certified within 40 minutes to one hour of interaction  
depending on the batch complexity. 
We issued for each batch a single instruction to close all the gaps by only using the MCP LPTP tools.
Table \ref{tab:mcp-benchmark} lists properties generated and certified for the main predicates. 

\begin{table}[htbp]
\centering
\begin{tabular}{|l|p{11.5cm}|}
\hline
\textbf{Program} & \textbf{Properties} \\
\hline
P14 (\texttt{dupli/2}) 
& $\forall l, m: \text{list}(l) \land \text{dupli}(l,m) \Rightarrow \text{list}(m)$ \newline
 $\forall l, m: \text{gr}(l) \land \text{dupli}(l, m) \Rightarrow \text{gr}(m)$ \newline
 $\forall l, m: \text{list}(l) \Rightarrow \textbf{terminates}\ \text{dupli}(l, m)$ \newline
 $\forall l, m_1, m_2: \text{dupli}(l, m_1) \land \text{dupli}(l, m_2) \Rightarrow m_1 = m_2$ \\
\hline
P15 (\texttt{dupli/3}) 
& $\forall l, n, m: \text{list}(l) \land \text{nat}(n) \land \text{dupli}(l, n, m) \Rightarrow \text{list}(m)$ \newline
 $\forall l, n, m: \text{gr}(l) \land \text{gr}(n) \land \text{dupli}(l, n, m) \Rightarrow \text{gr}(m)$ \newline
 $\forall l, n, m: \text{list}(l) \land \text{nat}(n) \Rightarrow \textbf{terminates}\ \text{dupli}(l, n, m)$ \newline
 $\forall l, n, m_1, m_2: \text{dupli}(l, n, m_1) \land \text{dupli}(l, n, m_2) \Rightarrow m_1 = m_2$ \\
\hline
P16 (\texttt{drop/3})
& $\forall l, n, m: \text{list}(l) \land \text{nat}(n) \land \text{drop}(l, n, m) \Rightarrow \text{list}(m)$ \newline
 $\forall l, n, m: \text{gr}(l) \land \text{nat}(n) \land \text{drop}(l, n, m) \Rightarrow \text{gr}(m)$ \newline
 $\forall l, n, m: \text{list}(l) \land \text{nat}(n) \Rightarrow \textbf{terminates}\ \text{drop}(l, n, m)$ \newline
 $\forall l, n, m_1, m_2: \text{list}(l) \land \text{nat}(n) \land \text{drop}(l, n, m_1) \land \text{drop}(l, n, m_2) \Rightarrow m_1 = m_2$ \\
\hline
P17 (\texttt{split/4})
& $\forall l, n, l_1, l_2: \text{list}(l) \land \text{nat}(n) \land \text{split}(l, n, l_1, l_2) \Rightarrow \text{list}(l_1) \land \text{list}(l_2)$ \newline
 $\forall l, n, l_1, l_2: \text{gr}(l) \land \text{split}(l, n, l_1, l_2) \Rightarrow \text{gr}(l_1) \land \text{gr}(l_2)$ \newline
 $\forall l, n, l_1, l_2, l_3, l_4: \text{split}(l, n, l_1, l_2) \land \text{split}(l, n, l_3, l_4) \Rightarrow l_1 = l_3 \land l_2 = l_4$ \\
\hline
P18 (\texttt{slice/4})
& $\forall l, i, k, r: \text{list}(l) \land \text{nat}(i) \land \text{nat}(k) \land \text{slice}(l, i, k, r) \Rightarrow \text{list}(r)$ \newline
 $\forall l, i, k, r: \text{gr}(l) \land \text{gr}(i) \land \text{gr}(k) \land \text{slice}(l, i, k, r) \Rightarrow \text{gr}(r)$ \newline
 $\forall l, i, k, r: \text{list}(l) \land \text{nat}(i) \land \text{nat}(k) \Rightarrow \textbf{terminates}\ \text{slice}(l, i, k, r)$ \newline
 $\forall l, i, k, r_1, r_2: \text{list}(l) \land \text{nat}(i) \land \text{nat}(k) \land \text{slice}(l, i, k, r_1) \land \text{slice}(l, i, k, r_2) \Rightarrow r_1 = r_2$ \\
\hline
P19 (\texttt{rotate/3})
& $\forall x, n, y: \text{list}(x) \land \text{nat}(n) \land \text{rotate}(x, n, y) \Rightarrow \text{list}(y)$ \newline
 $\forall x, n, y: \text{gr}(x) \land \text{gr}(n) \land \text{rotate}(x, n, y) \Rightarrow \text{gr}(y)$ \newline
 $\forall l, n, r_1, r_2: \text{list}(l) \land \text{nat}(n) \land \text{rotate}(l, n, r_1) \land \text{rotate}(l, n, r_2) \Rightarrow r_1 = r_2$ \\
\hline
P20 (\texttt{remove\_at/4})
& $\forall x, l, n, r: \text{list}(l) \land \text{nat}(n) \land \text{remove\_at}(x, l, n, r) \Rightarrow \text{list}(r)$ \newline
 $\forall x, l, n, r: \text{gr}(x) \land \text{gr}(l) \land \text{gr}(n) \land \text{remove\_at}(x, l, n, r) \Rightarrow \text{gr}(r)$ \newline
 $\forall x_1, x_2, l, n, r_1, r_2: \text{list}(l) \land \text{nat}(n) \land \text{remove\_at}(x_1, l, n, r_1) \land \newline \text{remove\_at}(x_2, l, n, r_2) \Rightarrow x_1 = x_2 \land r_1 = r_2$ \\
\hline
P21 (\texttt{insert\_at/4})
& $\forall x, l, n, r: \text{list}(l) \land \text{nat}(n) \land \text{insert\_at}(x, l, n, r) \Rightarrow \text{list}(r)$ \newline
 $\forall x, l, n, r: \text{gr}(x) \land \text{gr}(l) \land \text{gr}(n) \land \text{insert\_at}(x, l, n, r) \Rightarrow \text{gr}(r)$ \newline
 $\forall x, l, n, r_1, r_2: \text{list}(l) \land \text{nat}(n) \land \text{insert\_at}(x, l, n, r_1) \land \text{insert\_at}(x, l, n, r_2) \Rightarrow r_1 = r_2$ \\
\hline
P22 (\texttt{range/3})
& $\forall a, b, r: \text{nat}(a) \land \text{nat}(b) \land \text{range}(a, b, r) \Rightarrow \text{list}(r)$ \newline
 $\forall a, b, r: \text{gr}(a) \land \text{gr}(b) \land \text{range}(a, b, r) \Rightarrow \text{gr}(r)$ \newline
 $\forall a, b, r_1, r_2: \text{nat}(a) \land \text{nat}(b) \land \text{range}(a, b, r_1) \land \text{range}(a, b, r_2) \Rightarrow r_1 = r_2$ \\
\hline
P23 (\texttt{rnd\_select/3})
& $\forall l, n, r: \text{list}(l) \land \text{nat}(n) \land \text{rnd\_select}(l, n, r) \Rightarrow \text{list}(r)$ \newline
 $\forall l, n, r: \text{gr}(l) \land \text{gr}(n) \land \text{rnd\_select}(l, n, r) \Rightarrow \text{gr}(r)$ \\
\hline
P24 (\texttt{lotto/3})
& $\forall n, m, r: \text{nat}(n) \land \text{nat}(m) \land \text{lotto}(n, m, r) \Rightarrow \text{list}(r)$ \newline
 $\forall n, m, r: \text{gr}(n) \land \text{gr}(m) \land \text{lotto}(n, m, r) \Rightarrow \text{gr}(r)$ \\
\hline
\end{tabular}
\caption{Properties generated and certified via MCP for P14-P24}
\label{tab:mcp-benchmark}
\end{table}

\section{Related work}
\label{Related-work}
The recent arrival and rapid adoption of LLM's and their growing aptitude at writing and/or dealing with code has triggered different AI assisted programming practices or methodologies. We can distinguish the following, even though in practice they tend to overlap and hybrid approaches exist~\cite{10.1145/3759425.3763390,sapkota2025vibevsagentic,wang2026vibecontract}. 

Among these methodologies, \textit{vibe-coding}~\cite{meske2025vibereconfig, sarkar2025vibecoding} is presumably the most widespread and most easily accessible technique, for novices as well as experienced professionals. Vibe-coding is mostly understood as the process in which the user provides the LLM with a prompt representing an informal specification of a problem written in natural language, the LLM produces a solution in code, and the user tests and ideally accepts (possibly after multiple iterations) the given solution~\cite{fawzy2025vibepractice, sarkar2025vibecoding}. While the burden of validation lies with the user, preliminary reviews in the literature argue that testing is often skipped or even delegated to the AI~\cite{fawzy2025vibepractice}. Consequently, benchmarks show that code generated by vibe-coding often exhibits functional correctness issues and sometimes raises serious security concerns~\cite{fawzy2025vibepractice, zhao2025vibesafe}. 

A natural approach to alleviate some of the issues presented by vibe-coding is to augment the process with formal specification and verification (see, a.o., \cite{10.1145/3759425.3763390,wang2026vibecontract}). 
In what is sometimes called \textit{vericoding}~\cite{DBLP:journals/corr/abs-2509-22908}, the user provides a \textit{formal} specification of the problem upfront, thereby limiting accessibility of the technique to a more expert audience. The LLM then generates a verifiable implementation of the specification, i.e. it accompanies the generated code with a machine-checkable proof of its correctness. While the use of formal verification can in principle eliminate the functional correctness issues sometimes encountered in vibe-coding~\cite{DBLP:journals/corr/abs-2509-22908,10.1145/3759425.3763390}, correctness (and other) guarantees are obviously highly dependent on the quality and completeness of the specification.
Available benchmarks~\cite{DBLP:journals/corr/abs-2509-22908, algoveri2026} show results depending on the formal language used. In these benchmarks, using Dafny seems to perform better than Verus and Lean~\cite{DBLP:journals/corr/abs-2509-22908, algoveri2026}, but the effective success rates vary widely and, as the authors note, are subject to the rapid progress of the LLM models themselves. While these results are encouraging, other results are less optimistic. For example, in~\cite{thakur2025clever}, the authors report very low success rates when evaluating a more demanding benchmark in which both the specification and the implementation are required to be generated and verified via Lean.

While in vibe-coding and vericoding a human user orchestrates the code (and, in the case of vericoding, the proof) generation, \textit{agentic coding}~\cite{hassan2025agenticroadmap, wang2025agenticsurvey} goes further in the sense that the human user formulates a higher-level goal, and delegates to an autonomous AI agent a large part of the software production process (coding, developing and running tests, fixing bugs, and up to submitting pull requests) without human intervention. Verification is mostly empirical. While some speedups are reported, these appear highly context and task dependent~\cite{agarwal2026aiides, becker2025metrproductivity}. 
Moreover, the use of agents also appears to induce persistent quality risks. A recent analysis~\cite{metr2026swebench} reports that roughly half of agentic pull requests (for code passing automated tests) were rejected by the maintainers of a repository.

\section{Conclusion}
\label{Conclusion}

In this paper, we report on an experiment in which the first third of the well-known P-99 set of Prolog exercises is solved using LPTP in combination with an LLM.
We used Claude (Anthropic) in Code mode with the Opus 4.6 model.
By solving an exercise, we mean reading the natural language specification,
providing the Prolog code, constructing and running a test file, 
and stating and proving the usual properties of the Prolog code
(types, groundness, termination, existence, uniqueness).
For functional properties, we interacted with the LLM and checked each output.
The time needed to solve an exercise varies
from 15 minutes (P01) to several hours (P35). We note that the code and proofs of
P35 define an LP (logic programming) view of the 
\emph{prime factorization theorem} of arithmetic.
The final part of the paper presents ongoing work aimed at supporting lower-level interaction through the Model Context Protocol, enabling connections to any LLM that supports the protocol (\eg Gemini 3.1 Pro).

The \emph{vibe-coding} part of the experiment—reading the specification, and writing the Prolog code and tests—was handled easily by the LLM.
We first checked these outputs to ensure that we were working with the correct code, free from impure constructs such as the cut and from the use of Prolog’s native numbers.

The \emph{vericoding} part of the experiment—stating and proving reliability guarantees of the generated code—was clearly more challenging for the LLM; however, it was generally able to handle it, except for functional properties.
One should keep in mind that the properties 
proved by the LLM have been the subject of decades
of research work in program
verification and abstract interpretation \cite{Cousot77a}, 
in particular in the LP setting \cite{Cousot92a}.
We have now many theoretical results and some
implementations (\eg \cite{HermenegildoPBL05,PayetM06}) 
which compute more precise results than those inferred by the LLM.
We have already begun investigating the use of automated theorem provers to verify properties of Prolog programs expressed in LPTP \cite{MesnardMP26}, as well as the automatic generation of LPTP proofs for groundness properties \cite{MarianneMP25}.

Nonetheless, the ease and rapidity with which the LLM was able to grasp the (largely unknown) LPTP formalism are impressive.
We even observed that, when faced with a difficult proof, Claude examined the Prolog source code of the proof checker to gain a deeper ``understanding'' of the proof-checking process and fix the problematic proof.
Also combining an LLM with a proof checker avoids the ``hallucinations'' that the LLM may produce and allows it to fix its errors.
Thus, the LPTP/LLM combination appears to be a valuable tool for the LP developer. This applies both to constructing LPTP proofs of invariants inferred by abstract interpretation—when no purely algorithmic method is available—and to proving functional properties of Prolog code, where the corresponding abstract domain is typically not implemented as it depends on the specific problem (see Section~\ref{Selected-examples}).
An interesting problem to investigate is the \emph{scalability} of this approach.

\vspace{0.2cm}
\noindent
\textbf{Acknowledgement.} We thank Romain Pabot for enlightening discussions about LLMs.

\bibliographystyle{eptcs}

\end{document}